\documentclass{emulateapj}

\newcommand{\arcs}{\mbox{\ensuremath{^{\prime\prime}}}}

\usepackage{natbib}
\usepackage{longtable,lscape,psfig}
\shorttitle{Stellar X-ray sources in COSMOS}
\shortauthors{Wright et al.}

\begin{document}

\title{Stellar X-ray sources in the {\it Chandra} COSMOS survey}

\author{N.J.~Wright, J.J.~Drake and F.~Civano}
\affil{Center for Astrophysics, 60 Garden Street, Cambridge, MA~02138}
\email{nwright@cfa.harvard.edu}

\begin{abstract}

We present an analysis of the X-ray properties of a sample of solar- and late-type field stars identified in the {\it Chandra} Cosmic Evolution Survey (COSMOS), a deep (160~ks) and wide ($\sim$0.9~deg$^2$) extragalactic survey. The sample of 60 sources was identified using both morphological and photometric star/galaxy separation methods. We determine X-ray count rates, extract spectra and light curves and perform spectral fits to determine fluxes and plasma temperatures. Complementary optical and near-IR photometry is also presented and combined with spectroscopy for 48 of the sources to determine spectral types and distances for the sample. We find distances ranging from 30~pc to $\sim$12~kpc, including a number of the most distant and highly active stellar X-ray sources ever detected. This stellar sample extends the known coverage of the $L_X$-distance plane to greater distances and higher luminosities, but we do not detect as many intrinsically faint X-ray sources compared to previous surveys. Overall the sample is typically more luminous than the active Sun, representing the high-luminosity end of the disk and halo X-ray luminosity functions. The halo population appears to include both low-activity spectrally hard sources that may be emitting through thermal bremsstrahlung, as well as a number of highly active sources in close binaries.

\end{abstract}


\section{Introduction}

Nearly all types of star are known to emit X-rays through a range of different emission mechanisms that include shocks in the radiatively-driven winds of massive stars and emission from high-temperature stellar coronae in later-type stars \citep{vaia81}. Across the stellar sequence the level of X-ray emission varies from $L_X / L_{bol} \sim 10^{-8} - 10^{-3}$, but can also vary over several orders of magnitude within each spectral class. Close or interacting late-type binary systems, as well as active young single stars may even emit at much higher levels for short periods of time through flare events. Only evolved late-type giants and main sequence B and A-type stars do not appear to emit X-rays at these levels and their X-ray emission properties, if they emit X-rays at all, are still unknown \citep[e.g.][]{schm97}.

Solar- and late-type stars such as our Sun emit X-rays from a magnetically-confined plasma at typical temperatures of one to several million Kelvin known as a corona. The corona is thought to be heated mostly by magnetic reconnection events, powered by the stellar dynamo, which itself is thought to be generated -- at least in the Sun -- by differential rotation between the star's radiative and convective layers \citep[e.g.][]{skum72,pall81,noye84}. The observed decrease in stellar X-ray luminosity of several orders of magnitude between the zero age main sequence \citep[e.g.][]{feig02,flac06,wrig10a} and solar age \citep[e.g.][]{pere00} has therefore been attributed to the rotational spin-down of the star, though a consistent picture has yet to emerge. \citet{gude97} studied a sample of nearby solar-type stars aged $1-10$~Gyr and found that the X-ray luminosities decayed as $L_X \propto t^{-1.5}$, while \citet{mice02} could find no evidence for a clear decay law over a similar age range, and \citet{feig04} estimated a decay law of $L_X \propto t^{-2}$ from a sample of faint high Galactic latitude main-sequence stars.

The origin of these discrepancies could lie with the small but diverse samples used to study coronal X-ray emission. Wide-field X-ray surveys from the {\it Einstein} and {\it ROSAT} observatories \citep[e.g.][]{gioi84,voge99} have resulted in large samples of stellar X-ray sources \citep[e.g.][]{huns99,schm04,ague09}, but which are biased toward bright and nearby ($< 100$~pc) thin-disk stars. The detection of X-rays from more distant, low-metallicity stars in the Galactic halo is important because it allows us to study how X-ray activity behaves at low metallicity as well as probe the close binary population of the early Galaxy. Deep surveys with the {\it Chandra} and {\it XMM-Newton} observatories are necessary for detecting these distant sources \citep[e.g.][]{feig04,cove08}. With this aim in mind, we are mining multiple {\it Chandra} datasets to build large samples of X-ray sources that fully populate the stellar X-ray luminosity function. This will be necessary if we are to develop a full understanding of the influences of age, spectral type, and metallicity on stellar X-ray emission.

The Cosmic Evolution Survey \citep[COSMOS,][]{scov07} is a deep and wide extragalactic survey designed to probe the medium redshift galaxy and active galactic nuclei (AGN) populations. The COSMOS field has been observed at nearly all wavelengths with both ground- and space-based facilities \citep[e.g.][]{capa07,scov07b} providing a large multi-wavelength catalog for studies of galaxy evolution. The survey also samples a long sight-line through the Galactic disk and halo that may be used to study the stellar populations in these regions \citep[e.g.][]{robi07} and their properties at different wavelengths.

In this paper we present a sample of stellar X-ray sources identified in the {\it Chandra} survey of the COSMOS field \citep{elvi09}, including an analysis of their properties. In Section~2 we describe the identifications of the stellar X-ray sample and present complementary optical spectroscopy to confirm the stellar nature of the majority of sources, determine spectral types and distances and compare our sample with other stellar X-ray surveys. In Section~3 we analyze the coronal properties of these sources in the context of other stellar coronal studies.

\section{The {\it Chandra} COSMOS stellar sample}

The X-ray observations presented here are from the {\it Chandra} COSMOS Survey \citep{elvi09} that has imaged an area of $\sim$0.9~deg$^2$ of the COSMOS field using the ACIS\footnote{Advanced CCD Imaging Spectrometer} imager \citep{garm03} on board the {\it Chandra} X-ray Observatory \citep{weis02}. The survey uses a grid of 36 overlapping pointings to give a highly uniform exposure of $\sim$160~ks over the central 0.5~deg$^2$ and $\sim$80~ks over an outer region of 0.4~deg$^2$. A detailed source detection procedure \citep{pucc09} resulted in a catalog of 1761 sources detected in one or more X-ray bands, with well-defined sensitivities and completeness fractions as a function of both X-ray band and survey area.

\citet{civa10} used optical and near-IR observations of the COSMOS field to make identifications for 1750 of the 1761 {\it Chandra} sources, including 61 stars. Identifications were made using likelihood ratio tests and by comparing optical and near-IR images with the morphology of the X-ray source. 27 of the stars were identified morphologically or through positional alignment of an X-ray source with a bright star. A further 21 stellar identifications were made by fitting multi-wavelength photometry to the spectral energy distributions (SEDs) of templates taken from \citet{salv09}, while the remaining 13 were identified spectroscopically from dedicated COSMOS spectroscopic campaigns (see Section~\ref{s-spectra} for more details). In many cases these sources were identified as stars by multiple methods (e.g. photometrically identified and then confirmed spectroscopically). The majority of the remaining {\it Chandra}-COSMOS sources were identified as galaxies based on the above methods, with only 11 sources remaining unidentified by \citet{civa10}: 2 of these have no identifiable counterpart and 9 either have multiple possible counterparts or are either associated with faint optical sources in close angular proximity to bright stars or galaxies, such that their properties cannot be studied. Statistically, based on the 1750 identifications made, of which only 3.5\% are stars, the 11 unidentified sources are likely to be galaxies.

\subsection{X-ray photon extraction and spectral fitting}

Since the extraction and characterization of {\it Chandra} COSMOS sources performed by the survey collaboration has been conducted on the basis that they are extragalactic sources (which have different morphological and spectral properties to stellar sources) we have re-analyzed the observations of the stellar sources assuming that they are stellar. This was done using CIAO\footnote{Chandra Interactive Analysis of Observations, http://cxc.harvard.edu/ciao}~4.2 \citep{frus06}, CALDB~4.2.2, and the ACIS~{\sc Extract}\footnote{http://www.astro.psu.edu/xray/docs/TARA/ae\_users\_guide.html} code \citep[AE,][]{broo02} using the method outlined in \citet{wrig09a}. To summarize, AE uses a variety of point spread functions (PSFs) appropriate for the off-axis angle of each observation of each source to extract photons in a set fraction of the PSF (typically 90\%). The background is estimated from a region surrounding this PSF that excludes the PSFs of other sources. From these extractions, AE calculates a source significance and the Poisson probability, $P_{not}$, that the source counts are a superposition of background photons. At this point all sources were inspected visually and compared to the position of their designated optical counterpart to confirm their association. We then applied a cut to the sample, discarding any sources that had a higher probability of being a false source than of being a real source (i.e. $P_{not} > 0.5$), which resulted in one source being discarded and reduced our sample to 60 sources. This rather liberal cut level was chosen to maintain a high level of completeness with respect to the existing {\it Chandra} COSMOS catalog.

Spectral fitting was performed for the 27 sources with $>20$ net counts using {\sc xspec}\footnote{http://heasarc.nasa.gov/docs/xanadu/xspec} version~12.6.0 \citep{arna96}. The spectra were compared to {\sc apec} \citep{smit01} spectra corresponding to single-temperature thermal plasma models \citep{raym77} in collisional ionization equilibrium and absorbed by a hydrogen column density using the {\sc tbabs}\footnote{http://astro.uni-tuebingen.de/nh} {\sc xspec} model \citep{balu92}. Due to the low Galactic extinction in the COSMOS sight-line the hydrogen column density was allowed to vary only up to the maximum value for the field of $\sim 2 \times 10^{20}$~cm$^{-2}$ \citep{kalb05}, while the thermal plasma temperature was allowed to vary freely. A grid of initial thermal plasma temperatures covering $kT = 0.7-2.6$ was used to prevent fitting local minima and the model with the lowest C-statistic \citep{cash79} was then used for each source. Two-temperature thermal plasma models were also tested for these sources, but only the brightest source, CID\footnote{{\it Chandra} COSMOS ID number.}~546, had sufficient counts to produce a noticeably better fit using a two-component model. For the 33 sources with less than 20 net counts we used the method outlined in \citet{getm10} to calculate X-ray fluxes from count rates and median photon energies. Since the hydrogen column density in our field of view (FoV) is negligible, we use apparent X-ray fluxes as intrinsic fluxes. Uncertainties on these fluxes are not specified individually, but were determined statistically by \citet{getm10}. They are an approximate function of the net counts of the source and range from 30\% for sources with $\sim$20 net counts to $> 70$\% for sources with $< 5$ net counts.

The X-ray properties of the 60 retained sources are listed in Table~\ref{xstars_cosmos}. With the exception of a single very bright source (CID~546), the majority of sources have count rates of $10^{-5} - 10^{-3}$, appropriate for a sample based on observations of 100-200~ks and a source detection procedure that has extracted sources down to $\sim$3 net counts.

\subsection{Optical and near-IR photometry and spectroscopy}
\label{s-spectra}

Optical and near-IR photometry was taken from the COSMOS optical catalogs\footnote{Website: 
http://cosmos.astro.caltech.edu/data/index.html} that include data from the Sloan Digital Sky Survey \citep[SDSS,][]{york00}, the Subaru photometric catalog \citep{capa07}, and the CFHT\footnote{Canada-France-Hawaii Telescope}/Megacam catalog \citep{mccr10}. For the brightest sources that saturate in the deep COSMOS catalogs we complemented this data with near-IR photometry from the Two Micron All Sky Survey \citep[2MASS,][]{cutr03}. The majority of sources in our stellar sample have either 2MASS or SDSS photometry and therefore for the faintest sources we used photometry in other photometric systems translated to the SDSS or 2MASS systems using the conversions listed in \citet{capa07}. All photometry is listed in Table~\ref{xstars_oir}.

Spectra for objects in our sample were compiled from a number of sources and used to determine spectral types and confirm the stellar nature of the sources. Where available, spectroscopy was taken from the accumulated spectroscopic catalogs available for the COSMOS field, including data from IMACS\footnote{Inamori Magellan Areal Camera and Spectrograph}/Magellan \citep{trum07,trum09} and VIMOS\footnote{VIsible MultiObject Spectrograph}/VLT\footnote{Very Large Telescope} \citep{lill07,lill09}. This amounted to 13 stellar spectra.

\begin{figure}
\begin{center}
\includegraphics[height=240pt, angle=270]{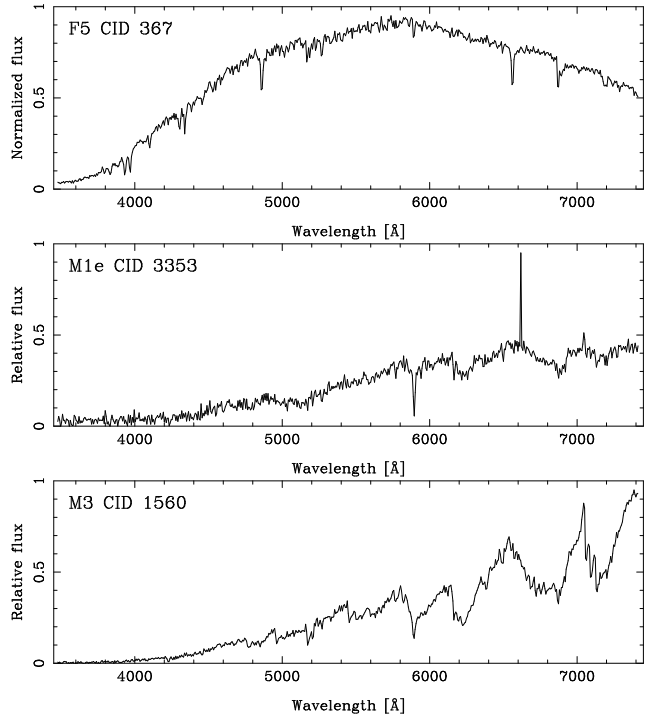}
\caption{Optical identification spectra of three sources observed with the FAST spectrograph on the 1.5-m Tillinghurst telescope. Spectra are not flux calibrated and are shown in normalized flux units.}
\label{spectra}
\end{center}
\end{figure}

An additional 35 spectra (for sources with $r \leq 17$) were obtained using the FAST spectrograph \citep{fabr98} on the 1.5-m Tillinghurst telescope at the Fred Lawrence Whipple Observatory, Mount Hopkins, Arizona. The spectrograph was equipped with a 300~gpm grating, resulting in a resolution of $\sim$3~\AA\ and a wavelength coverage of 3480--7400~\AA. Identification spectra were obtained from the raw data following standard data reduction procedures including bias-subtraction, flat-fielding, cosmic-ray removal and wavelength calibration. The exposure time per source ranged from 1 to 30 minutes, and the seeing was typically 1--2\arcs. Figure~\ref{spectra} shows examples of these spectra.

Spectral classifications for the 48 sources with spectroscopy were obtained by visual comparison with low-resolution spectra of MK standards, assuming that all stars lie on the main sequence\footnote{We have assumed that all these stars are on the main sequence since there are no known star forming regions along the line of sight, and evolved, late-type giants are known to be very weak X-ray emitters \citep[e.g.][]{lins79,ayre81} and therefore unlikely to be detected in our sample. It is possible that a giant might have an active secondary, and \citet{cove08} identify a number of confirmed and potential giant stars in their sample and find that 2-10\% of their sample are likely to be giants. However, because giants are more luminous than dwarfs they are typically detected at greater distances, and therefore the X-ray emission from an active secondary would have to be particularly high to be detectable at such a distance. Therefore, while it should be noted that a small fraction of our sources could be giants, for simplicity we will assume that they all lie on the main-sequence.}. For solar-type stars (F/G/K-type) we used spectra from \citet{gray09} obtained at the Dark Sky Observatory and available online\footnote{http://stellar.phys.appstate.edu/Standards/stdindex.html}, while for cooler stars (late K \& M-type) we used the atlas of late-type spectra presented by \citet{kirk91}. The spectral types determined range from F5 to M7 (see Figure~\ref{spectra} for examples), with some of the M-type stars showing H$\alpha$ in emission, suggesting chromospheric activity, which is a good tracer of coronal activity in late-type stars.

To estimate the uncertainty on our spectral types we classified each star independently 5 times. The mean spectral type (assigning integer values to each subtype) was taken as the final spectral type and the standard deviation was calculated for each star as an indicator of the uncertainty. For solar type stars we found the uncertainty to be approximately $\pm2$ subtypes, but only $\pm1$ subtype for late-type stars (except for objects where the spectral quality was poor and the uncertainty is $\pm2$ subtypes). All of the objects with spectra were confirmed as stellar sources, supporting the accuracy of the SED fitting technique used to separate stars from galaxies.

For the 12 sources without spectra we determined photometric spectral types based on their $g-i$ colors and the empirical colors from \citet{cove07}. A comparison of the photometric and spectroscopically-determined spectral types for those stars with both spectra and SDSS photometry shows an excellent agreement with a standard deviation of only 1.1 subtypes (a similar method using $J-K$ photometry and empirical colors from \citet{keny95}, \citet{cove07}, or \citet{krau07} is less accurate with a standard deviation of $\sim$3 subtypes). Based on this test we assign an uncertainty of $\pm2$ subtypes for photometrically-determined spectral types. The final distribution of spectral types determined from spectroscopy (photometry) is 6 (1) F-type stars, 7 (0) G-type stars, 19 (1) K-type stars, and 16 (10) M-type stars. The accumulated photometry and spectral types are listed in Table~\ref{xstars_oir}.

Finally, we searched through the SIMBAD\footnote{Website: http://simbad.u-strasbg.fr/simbad/} Astronomical data base for previously identified objects in our sample. Three sources have previously determined spectral types: CID~590 is listed as K0 (we find it to be a K2 type), CID~1560 is listed as M-type by \citet{wels07} (we find M3 type) and CID~3381 is listed by \citet{hein92} as a K0-type multiple star, while we find it to be G8 type. All of these differences are within our classification uncertainties. We also identified two matches (CIDs 537 and 1560) between our catalog and the catalog of flaring M-dwarf stars from the {\it Galactic Evolution Explorer} (GALEX) all-sky survey \citep{wels07}, a result that is not surprising since objects that flare in the ultraviolet are also likely to exhibit pronounced X-ray emission because of their flaring and therefore make them easier to detect in a flux-limited sample such as our own.

\subsection{Distances and X-ray luminosities}

Distances were determined for all sources based on the spectral types estimated above and using absolute magnitudes for main-sequence stars from \citet{krau07}. To accurately sample the most luminous part of the SED we use $r'$-band photometry for solar-type stars and $K$-band photometry for late-type stars (where both photometry exists we find that the mean difference between distances determined using these two methods is 18\%). We calculated uncertainties for these distances based on the uncertainties of the underlying spectral types and derive uncertainties of $\pm10$\% for solar-type stars and $\pm15$\% and $\pm30$\% for late-type stars with spectral types uncertain to $\pm1$ and $\pm2$ subtypes, respectively.

These distances are listed in Table~\ref{xstars_oir} and vary from $\sim$30~pc to over 10~kpc (see discussion of the most distant sources in Section~\ref{s-xstar1600}) with the majority of sources at distances of 100 -- 1000 pc. At the Galactic latitude of the COSMOS field, +42$^\circ$, a Galactic star counts simulation \citep{gira05} shows that the dominant stellar population switches from the disk to the halo at a distance of $\sim$400~pc. While this value may be slightly greater for an X-ray selected sample (because the halo population is older and will therefore have typically lower X-ray luminosities), and it is impossible to separate disk and halo stars individually based on their distances alone, this does suggest that a number of sources in our sample probably belong to the Galactic halo, though radial velocities or metallicity measurements will be necessary to be certain. Based on our comparison with other samples, some of these sources appear to be the most distant X-ray emitting late-type stars currently known.

\begin{figure*}
\begin{center}
\includegraphics[height=450pt, angle=270]{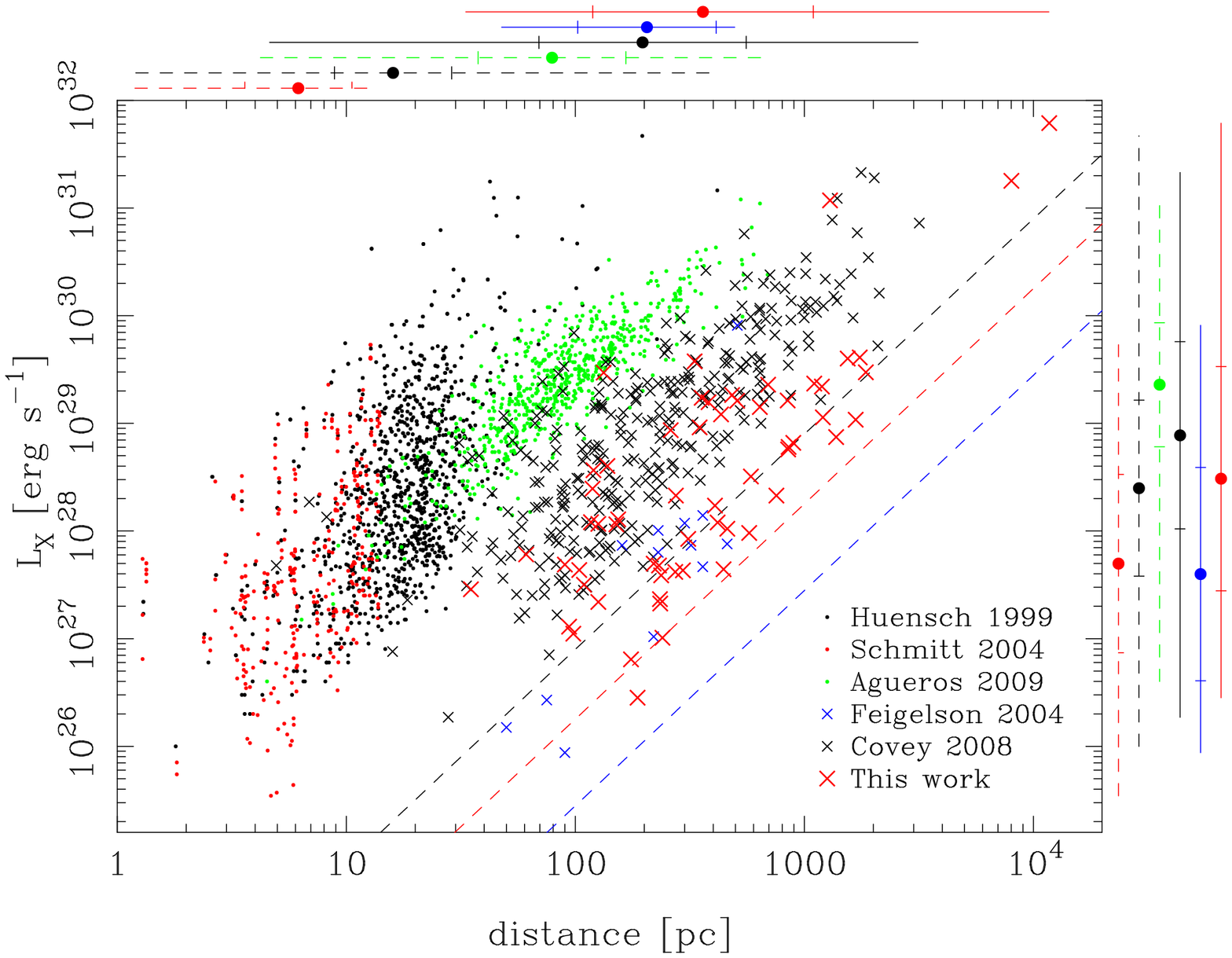}
\caption{X-ray luminosity as a function of distance for sources detected in this study (red crosses) and compared to recent samples. Samples shown are those of \citet{huen99} (black dots), \citet{schm04} (red dots), \citet{feig04} (blue crosses), \citet{ague09} (green dots), and \citet{cove08} (black crosses). Sources from {\it Chandra} studies are shown as crosses while those from ROSAT studies are shown as dots. For reference we also show dashed lines illustrating the estimated {\it Chandra} sensitivity limits using the formula provided by \citet{feig05} for the three {\it Chandra} surveys in the same color as their respective symbols. We also show the range of distances (above the figure) and X-ray luminosities (to the right) spanned by each sample as illustrated by dashed (for ROSAT) and full lines (for {\it Chandra}) in the same color as their respective symbols. Marked on each of these lines are the mean logarithmic values (shown with a filled circle) and the standard deviations (shown with two marks either side of the mean).}
\label{distances}
\end{center}
\end{figure*}

Distances were used to calculate X-ray luminosities for each source, which have uncertainties that scale as the square of the distance uncertainties, i.e. $\pm20$\% for solar-type stars, and $\pm30$\% and $\pm70$\% for the late-type stars. Figure~\ref{distances} shows the distribution of our sample in $L_X$--distance space compared to other recent surveys of stellar X-ray sources from {\it Chandra} and {\it ROSAT}. Also shown are the range of distances and X-ray luminosities spanned by each sample, as well as their logarithmic means and standard deviations. We also show estimates of the catalog sensitivity limits using the formula provided by \citet{feig05} for the three {\it Chandra} surveys, which shows that increases in survey depth should allow fainter and more distant sources to be detected. However, the transition from the shallow {\it ROSAT} surveys through the different {\it Chandra} surveys has been most apparent in an increase in source distance, with an actual shift toward {\it higher} X-ray luminosity. This bias is caused by the selection of fields for extragalactic observations (in the case of the CDF-N and COSMOS surveys) that avoid nearby bright stars and the exclusion of fields with bright sources in the {\it Chandra} Multiwavelength Project (ChaMP) survey \citep[as noted by][in their version of this figure]{cove08}.

\subsection{Comparison of {\it Chandra} surveys}

Limiting our comparison of X-ray luminosities to only {\it Chandra} surveys \citep[see][for a discussion of the limits of the {\it ROSAT} stellar samples]{cove08}, we note that our sample is similar to that of \citet{cove08}, but both are typically more X-ray luminous than the sample of \citet{feig04}, where all but one source have log~$L_X \lesssim 28$. To determine if these three samples are drawn from the same distribution we performed two-sample Kolmogorov-Smirnov (K-S) tests on each pair of surveys. Comparing our sample with that of \citet{cove08} we find a $\sim$10\% probability that they are drawn from the same distribution, a discrepancy caused by the lack of objects in our sample with log~$L_X = 30 - 31$. This may be partly due to size-of-sample effects, but it could also be caused by a change in the main population being sampled as we go from the thin-disk to the halo. The sources `missing' from our sample would be expected to lie at distances of $\sim 500 - 5000$~pc, a range dominated by halo stars that, due to their age, might not reach the levels of X-ray luminosity seen in younger, more local samples. The presence of the distant and highly luminous X-ray sources in our sample would then not be considered part of the single-star X-ray luminosity function but potentially due to rarer, active binary systems.

Performing the K-S test on the sample of \citet{feig04} we find a probability of $<$0.1\% that it is drawn from the same distribution of either of the other two {\it Chandra} samples. In the sample of \citet{feig04}, the first and third most X-ray faint sources were identified only in the full 2~Ms of {\it Chandra} Deep Field North (C-DFN) observations and not in the 1~Ms sample of 11 stars that the authors identify as the more uniform and complete subset of stellar emitters. Removing these and repeating the K-S test gives a probability of $\sim$5\% that the sample is drawn from the same distribution as ours. This discrepancy, which may simply be due to non-uniformities in the stellar distributions, is problematic because it opens up the question of whether surveys such as these can be used to study stellar X-ray emission. However, there are a number of differences in the depths, areas, and sensitivities of the three surveys that could present possible explanations for these differences. For example, the lack of any distant or highly X-ray luminous sources in the C-DFN sample could be attributed to a size-of-sample effect that might limit the most distant source detected in the smaller field of view (FoV) of the C-DFN survey compared to that of COSMOS.

Our lack of intrinsically faint X-ray sources is more difficult to understand, but we suggest this could be explained by an exposure time bias when calculating X-ray luminosities from time-averaged X-ray fluxes, as was done in both works. If a source can only be detected during a short-duration flare event and only flares once during the observation \citep[which is not unlikely, since][detect one flare every $\sim$2.4~Ms in the C-DFN observations]{feig04} then its X-ray luminosity will scale inversely with the total exposure time. This could explain why the sample of \citet{feig04} contains sources $\sim$10 times fainter in X-rays than our sample, because the typical observation time of the C-DFN stars is $\sim$10 times greater than those in the COSMOS survey. However, we note from studying the light curves shown by \citet{feig04} that many of the flaring sources detected by them appear to have quiescent X-ray luminosities sufficient for them to be detected even if they had not flared. This would therefore make this explanation unlikely.

As it is we are unable to identify the exact causes of the difference in X-ray luminosities between the two samples. We have searched the COSMOS field for X-ray emission from the nearest stellar sources in our FoV and find no evidence for X-ray emission from them. Additionally, we do not believe that any distant and highly luminous X-ray sources in the C-DFN were missed by \citet{feig04} since the optical photometry available to them was comparable in depth to ours, and the majority of C-DFN sources were observed spectroscopically, which would have allowed them to identify any faint stars. To achieve a K-S probability of $>$20\% that the two samples are drawn from the same distribution would require either $\geq 7$ high luminosity sources to have been missed from the C-DFN sample, or for $\geq 4$ existing sources to have notably higher X-ray luminosities. We can therefore only suggest that the differences in the two populations are a combination of the above reasons and the effects of Poisson statistics on such small samples.

\section{Discussion of source properties}

We now consider the X-ray properties of our sample of 60 stellar X-ray sources and explore correlations between their X-ray and optical properties. We also compare their properties to the recent {\it Chandra} surveys from \citet{feig04} and \citet{cove08}. As shown in Figure~\ref{distances} the former survey has sampled moderately distant (50-500~pc) intrinsically faint X-ray emitters, while the latter survey has probed more distant, but more luminous sources. Our sample represents an extension of the \citet{cove08} survey, including sources with a similar range of intrinsic X-ray luminosities but at greater distances and therefore likely including a number of halo sources.

\subsection{X-ray properties as a function of spectral type}

\begin{figure}
\begin{center}
\includegraphics[height=240pt, angle=270]{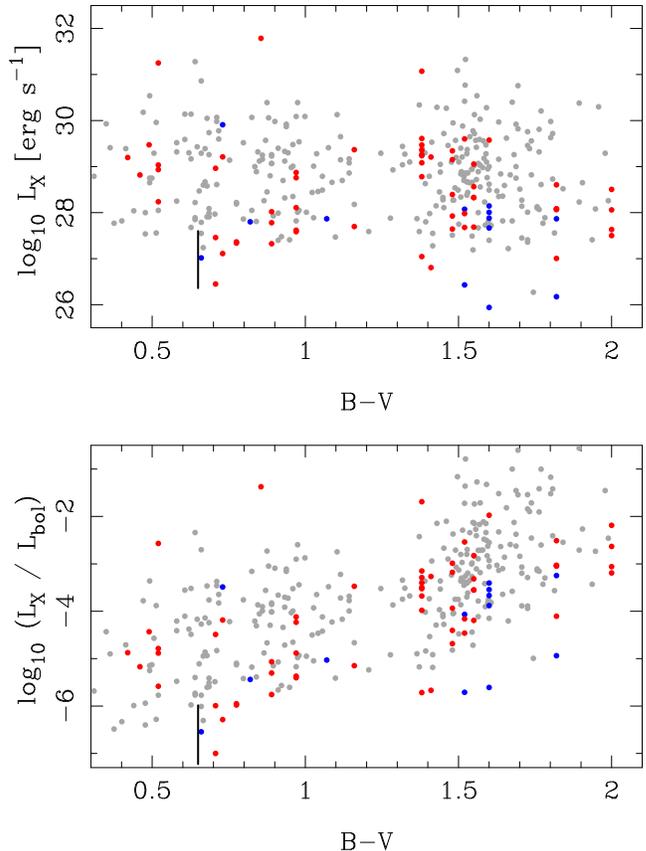}
\caption{X-ray luminosities (top panel) and X-ray to bolometric luminosity ratios (bottom panel) as a function of $B-V$ color for sources in our sample (red dots), the sample of \citet[][blue dots]{feig04}, and of \citet[][grey dots]{cove08}. The range of X-ray luminosities exhibited by the Sun \citep{pere00}, and adjusted to match the {\it Chandra} bands, is shown as a black line.}
\label{lx}
\end{center}
\end{figure}

We first consider the X-ray luminosities of our sample as a function of their spectral type, for which we use the proxy of $B-V$ color for comparison with previous works \citep[using the table of color as a function of spectral type presented by][]{keny95}, shown in Figure~\ref{lx}. Our sample includes sources with X-ray luminosities in the range log~$L_X \sim 27.5 - 30.5$~erg~s$^{-1}$, independent of spectral type. This differs from the finding of \citet{zick05} who found a weak trend of increasing $L_X$ toward earlier spectral types, though their nearby sample could differ intrinsically from our more distant sample, and may include fewer highly active stars. The X-ray luminosities are typically higher than that of the contemporary Sun \citep[][adjusted to match {\it Chandra}'s spectral bands]{pere00}, with a small number of G-type stars exhibiting X-ray luminosities similar to the active Sun. Figure~\ref{lx} also shows the X-ray to bolometric luminosity ratio as a function of spectral type, using bolometric luminosities determined from the table of main-sequence bolometric magnitudes presented by \citet{krau07}. Our sample and that of \citet{cove08} are in agreement with the observed trend of increasing $L_X / L_{bol}$ toward later spectral types \citep[e.g.][]{flem95}.

\subsection{X-ray properties of the most distant members of the sample}
\label{s-xstar1600}

Our sample includes a number of sources with distances $> 1$~kpc that are likely to be members of the Galactic halo. Partly because of our sensitivity limits, the majority of these are highly luminous with log~$L_X > 31$~erg~s$^{-1}$. Since the Galactic halo is $\sim$10~Gyrs old and $L_X$ declines with age through magnetic braking, these are almost certainly close binaries kept active through tidal interaction and tapping of orbital angular momentum to sustain strong dynamo activity. \citet{ottm97} found that Population~{\sc ii} binaries typically have lower X-ray luminosities than more metal-rich systems, but do exhibit a high-luminosity tail with log~$L_X \sim 29 - 31$~erg~s$^{-1}$. We are unable to definitively identify halo members amongst our sample of moderately bright (log~$L_X \sim 28 - 29$~erg~s$^{-1}$) sources, but our detection of two distant and highly-luminous sources with $d \gtrsim 8$~kpc and log~$L_X > 31$~erg~s$^{-1}$ suggests that a high-luminosity tail for the halo binary distribution does exist and at a higher X-ray luminosity than found by \citet{ottm97}. If our sample is representative, a simple extrapolation suggests that the Galactic halo contains $\sim 10^5$ binaries with log~$L_X \geq 31$~erg~s$^{-1}$.

Our most distant and highly luminous source, CID~1600, is a relatively faint detection with only $\sim$5 net counts and a probability of being a background event of 0.026, though the {\it Chandra} COSMOS catalog lists it with $\sim$12 net counts \citep{pucc09}. Optical and near-IR photometry suggests it is a K1-type star (though no spectroscopy exists) that would put it at a distance of $\sim$11.7~kpc and give it an X-ray luminosity of $L_X \sim 5 \times 10^{31}$~erg~s$^{-1}$. While the X-ray source is $\sim$3$^{\prime\prime}$ from the optical counterpart, the source is $\sim$12$^\prime$ off-axis, and its PSF is therefore similarly-sized. There are no other suitable optical counterparts in either the deep Hubble Space Telescope observations of the field \citep[that extend down to $I_{AB} \sim 27$,][]{scov07b} or longer wavelength mid-IR observations. Considering this we are left with the choice that the X-ray source is either associated with the optical source, or is a background event. We note that the chance of an X-ray source being within 3$^{\prime\prime}$ of a source with $r' \lesssim 22$ (as the optical source does) is  0.027, making the probability that a random background fluctuation would be found in such a location to be $\sim 7 \times 10^{-4}$. Considering that the {\it Chandra} COSMOS catalog contains $\sim$1700 sources we might expect at least one spurious event such as this and this could be a possible candidate. If the source were real it is likely to be in an active binary that was observed to flare during the observations, an interpretation supported by the high median photon energy of the source ($\bar{E_X} \sim 3.8$~keV) and the fact that all but one of the detected photons came in the second of two similar-length exposures. Deep spectroscopy will be necessary to confirm the stellar nature of the source and detect evidence of its binary nature.

The second most distant and luminous source in our sample, CID~3205, is a more reliable detection with $\sim$12 net counts and a close association with an optical source whose photometry suggests it is either an F8-type star at a distance of $\sim$8~kpc or a much closer white dwarf (WD) or cataclysmic variable \citep[since the colors of these sources overlap in the SDSS system,][]{fan99}. If the source were an F8 star, then since the main sequence lifetime of such a star is $\sim$5~Gyrs it could not be a member of the Galactic halo, but is more likely to be a member of the thin disk that has been forced onto a highly elliptical orbit through some sort of close encounter. While the X-ray luminosity of the source is high, it is not unfeasible if the source is relatively young, though it could also be evidence for the source being an active binary. However, if the source were a WD it would likely be much closer and therefore have a lower X-ray luminosity. The USNO-B catalog \citep{mone03} lists a small proper motion for the source of 0.26$^{\prime\prime}$/yr, which would be consistent either with a nearby WD or with a distant main sequence star on an elliptical orbit and at a high velocity. We note that the $u-g$ color of 1.01 is redder than would be expected for a WD \citep{smol04}, but at the blue end of the colors of F-type main-sequence stars \citep{cove07}. Again, spectroscopy will be necessary to confirm the spectral type of the source and detect evidence for binarity.

\subsection{Coronal plasma temperatures and variability}

\begin{figure}
\begin{center}
\includegraphics[height=240pt, angle=270]{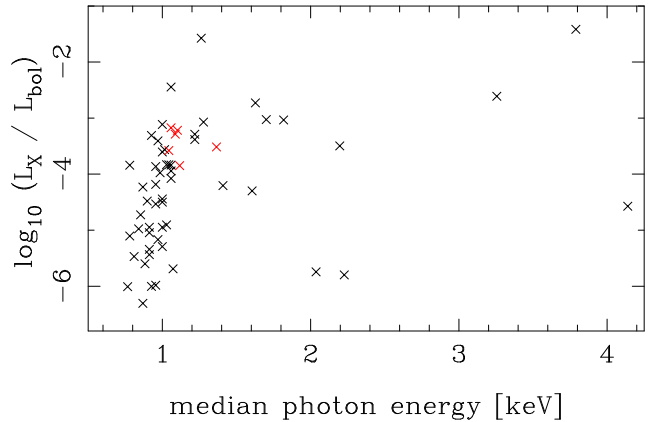}
\caption{X-ray to bolometric luminosity ratios as a function of median photon energies for all sources. Sources with identifiable flaring events are shown in red.}
\label{plasma}
\end{center}
\end{figure}

The median photon energy of X-ray source events provides a simple characterization of the X-ray emission properties that is particularly useful for faint sources lacking plasma temperatures derived from spectral fits. The distribution of median energies seen in Figure~\ref{plasma} is highly clustered around 1~keV and in agreement with the distribution of plasma temperatures from spectral fits, $\sim 0.5 - 1.0$~keV (6--12~MK), typical for stars of moderate to high activity levels and for active regions and flares on the Sun \citep{pere00}. This indicates that our sample includes a large fraction of active stars, as might be expected for a luminosity-limited sample. We also observe no correlation between either the plasma temperature determined from spectral fitting or the median photon energy of a source with its spectral type.

\begin{figure*}
\begin{center}
\includegraphics[height=470pt, angle=270]{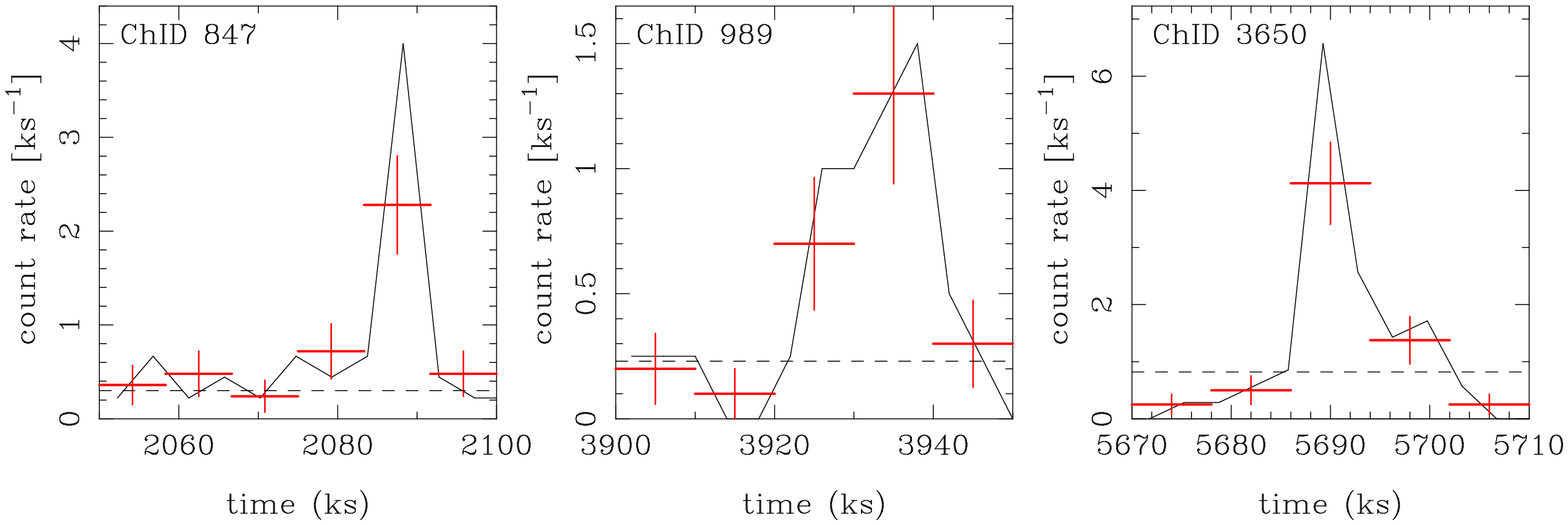}
\caption{Segments of the X-ray light curves for sources CID 847, 989, and 3650 (black lines) showing flaring events. Coarser bins are overlaid in red with Poisson uncertainties. The black dashed lines show the mean count rates over the total observations for each source.}
\label{lightcurves}
\end{center}
\end{figure*}

Figure~\ref{plasma} shows the X-ray to bolometric luminosity ratios as a function of median photon energy for all our sources. For sources with $\bar{E} \lesssim 1.5$~keV we note a trend of increasing luminosity ratio with median photon energy. This is similar to the relationship between the luminosity ratio and plasma temperature commonly seen in late-type stars and found by \citet{schr84} and \citet{schm90}. It is thought to result from the increasing size and intensity of active regions, and the growth of flaring activity as active regions fill larger fractions of the stellar surface \citep[e.g.][]{drak00}. In order to look for the influence of flares, we compiled X-ray light curves from the ACIS event lists and tested for variability. We used a one-sample Kolmogorov-Smirnov test to compare the distribution of photon arrival times with that expected for a constant source (the null hypothesis) and then derived the probability of accepting the null hypothesis, $P_{KS}$, as listed in Table~\ref{xstars_cosmos}. We studied the light curves for 17 sources with $P_{KS} < 0.01$ and identified six flaring events with durations of 2-5 hours, three of which are shown in Figure~\ref{lightcurves}. These six sources are also indicated in Figure~\ref{plasma} and can clearly be seen at the high $L_X / L_{bol}$ end of the trend mentioned above. We then studied the light curves of the other sources with high $L_X / L_{bol}$ and high $\bar{E}$, but could find no evidence for bright flares.

We find no trend of median photon energy with luminosity ratio for stars with higher median energies, corresponding to plasma temperatures of $\sim 15 - 45$~MK. The luminosity ratios for these stars range from $10^{-6} - 10^{-2}$. The majority of these sources are distant with 60\% of sources with $\bar{E} > 1.5$~keV found at distances $> 1$~kpc, compared to a fraction of 25\% for the entire sample. One explanation for the spectral hardness of sources with high X-ray luminosity ratios is that these sources were observed during particularly long and bright flares. However this cannot be the case for all the hard sources because they do not appear to be significantly more variable than the soft sources: 30\% of the hard sources have $P_{KS} < 0.01$, compared to 28\% for the entire sample, while all the clearly identified sources with flares are in the soft sample. There will be a bias in this analysis because variability is easier to identify in sources with more counts, which are more likely to be included in the nearby soft sample, but there appear to be multiple hard X-ray stellar sources that cannot be explained by variability. It is more likely that some of these sources, particularly those with low to moderate luminosity ratios (log~$L_X / L_{bol} < -4$), are members of the halo population that have extremely metal-poor coronae. Indeed, a lack of a correlation between X-ray luminosity and plasma temperature was noted by \citet{ottm97} based on a survey of nearby Population~{\sc ii} close binaries. These authors also found the Pop.~{\sc ii} stars to have harder spectra than their Pop.~{\sc i} counterparts, and attributed this to the lower radiative efficiency of metal-poor plasma. Our sample of stars appears to support this, with what must be halo stars appearing to have very hot coronae. In such coronae, it seems that the lack of plasma radiative cooling through metal lines is compensated for by much higher plasma temperatures and that radiative cooling occurs predominantly through the bremsstrahlung continuum.

Finally we note that CID~546 is the only source with sufficient counts to make a reasonable two-temperature thermal plasma fit. The difference between single and double-component thermal plasma fits is a decrease in the Cash statistic \citep{cash79} from 843 to 752 and a factor three drop in the maximum fit residuals from 0.015 to 0.005. The two-temperature thermal plasma fit consists of plasma at temperatures of 0.73 and 2.1~keV at a flux ratio of 1.3:1 \citep[e.g.][]{lope07}. Extraction of the light curve of CID~546 does not reveal any large flaring events, and the median photon energy remained relatively constant at $\sim$1.1~keV throughout the observations.

\section{Conclusions}

In this paper we have studied the stellar content of the {\it Chandra}-COSMOS survey and identified a sample of 60 stellar sources for which we present X-ray properties, as well as optical and near-IR photometry. In addition we have obtained spectroscopic classifications for 48 of the sources, confirming their stellar nature and allowing us to derive spectral types and distances. In the $L_X$-distance plane the sample extends the recent survey of \citet{cove08} to more distant sources, with the majority of sources lying at several hundred parsecs. The most distant sources are highly likely to be members of the Galactic halo, with the two most distant sources at $\sim$8--12~kpc being the most distant late-type stellar X-ray sources known.

The X-ray luminosity distribution of our sample is in approximate agreement with that of \citet{cove08}, but is significantly more luminous that that of \citet{feig04} and we consider a number of possible explanations for this, including an exposure-time bias and size-of-sample effects. Differences between high Galactic latitude stellar X-ray samples is potentially problematic because it raises the issue for how much can be learnt about stellar X-ray emission and dynamo activity from studies of individual sight lines. In a future paper we will attempt to model these samples and therefore explore the possibility that these differences are due to either size-of-sample effects or differences in the populations sampled that would be caused by different survey depths. Further studies in other equally deep {\it Chandra} surveys will also be useful to probe this discrepancy as well as increasing the overall sample size.

A comparison of X-ray and optical properties reveals no major differences between this and previous samples, though we note a large number of sources with high plasma temperatures and we suggest these are a combination of low-metallicity halo stars emitting through thermal bremsstrahlung at high temperatures, and a population of flaring close binaries in the Galactic halo. This reveals the excellent opportunity presented by this sample and other deep {\it Chandra} surveys to understand X-ray emission at low metallicities and probe the close binary population of the early Galaxy. High-resolution optical spectra will be necessary to measure metallicities and identify binaries.

A future paper will use this catalog to test models of the decay of X-ray activity in solar- and late-type stars in the Galaxy.

\acknowledgments

We would like to thank the referee, Kevin Covey, and the scientific editor, Eric Feigelson, for useful comments that improved the work presented here. We are grateful to the staff at the Fred Lawrence Whipple Observatory, particularly Perry Berlind and Michael Calkins, for FAST spectroscopy with the 1.5-m on Mt Hopkins, and to Susan Tokarz and Nathalie Marthinbeau for data reduction. This research has made use of data from the {\it Chandra X-ray Observatory} and software provided by the {\it Chandra X-ray Center} (CXC) in the application packages CIAO and Sherpa, and from Penn State for the {\sc ACIS Extract} software package. This work has made use of data from the {\it Chandra} COSMOS Survey, which is supported in part by NASA {\it Chandra} grant number GO7-8136A. We thank the C-COSMOS team for their work on this survey, assistance with this research, and careful reading of this manuscript. Special thanks are given to Tom Aldcroft, Marcella Brusa, Martin Elvis, Michael Rich, and Gianni Zamorani.  This research has also made use of NASA's Astrophysics Data System and the Simbad and VizieR databases, operated at CDS, Strasbourg, France. This work was funded by {\it Chandra} grant AR9-0003X. JJD was supported by NASA contract NAS8-39073 to the {\it Chandra} X-ray Center. 

{\it Facilities:} \facility{CXO}, \facility{FLWO}, \facility{2MASS}

\bibliography{/Users/nwright/Documents/Work/tex_papers/bibliography}

\tiny 
\begin{landscape}
\begin{longtable}{ccc ccc ccccc ccccc}
\caption[]{X-ray properties of stellar sources detected in the {\it Chandra} COSMOS survey.}\\ 
\tableline 
RA & Dec & CID & Cnts & \multicolumn{2}{c}{$\delta_{Cnts}$} & Sig. & log($P_{not}$) & Exp. & $\bar{E_X}$ & log($P_{KS}$) & Model fit & $kT$ & \multicolumn{3}{c}{X-ray fluxes (erg cm $^{-2}$ s$^{-1}$)} \\ 
\cline{5-6} \cline{14-16} 
 (J2000) & (J2000) & & (net) & Upper & Lower & ($\sigma$) & & (ks) & & (keV) & & (keV) & log $F$ & log $F_s$ & log $F_h$ \\ 
\tableline 
\endhead 
\tableline 
\multicolumn{16}{r}{\emph{continued on next page}}\\ 
\endfoot 
\tableline 
\endlastfoot 
10:00:48.44 & 2:07:34.8 & 48 &  144.67 & 13.3 & 12.2 &  10.9 & -6.00 & 187.88 & 1.00 & -0.20 &  1T & $0.74^{}_{}$ &  $-14.20^{+0.04}_{-0.04}$ & -15.34 & -14.17 \\ 
10:00:49.51 & 2:07:14.6 & 49 &  37.77 & 7.6 & 6.5 &  5.0 & -6.00 & 187.88 & 0.93 & -0.02 &  1T & $0.22^{+0.53}_{-0.11}$ &  $-13.75^{+0.08}_{-0.08}$ & -17.09 & -13.75 \\ 
10:00:20.96 & 1:44:32.3 & 268 &  30.96 & 6.9 & 5.8 &  4.5 & -6.00 & 96.19 & 1.03 & -0.75 &  1T & $0.51^{+0.78}_{-0.22}$ &  $-14.13^{+0.09}_{-0.09}$ & -15.76 & -14.12 \\ 
9:58:23.06 & 2:13:11.9 & 321 &  87.36 & 11.3 & 10.3 &  7.7 & -6.00 & 91.46 & 0.91 & -0.87 &  1T & $0.54^{}_{}$ &  $-13.97^{+0.05}_{-0.05}$ & -15.53 & -13.96 \\ 
10:00:57.46 & 1:55:48.7 & 367 &  157.32 & 14.0 & 13.0 &  11.2 & -6.00 & 184.36 & 0.84 & -0.40 &  1T & $0.40^{}_{}$ &  $-14.05^{+0.04}_{-0.04}$ & -16.05 & -14.04 \\ 
10:01:42.18 & 1:53:19.7 & 397 &  16.63 & 5.6 & 4.4 &  3.0 & -6.00 & 93.25 & 0.77 & -0.14 &  \ldots & \dots &  $-14.94^{}_{}$ & \ldots & \ldots \\ 
9:59:54.70 & 2:17:06.0 & 444 &  342.87 & 20.2 & 19.2 &  17.0 & -6.00 & 232.80 & 0.98 & -1.72 &  1T & $0.66^{}_{}$ &  $-13.84^{+0.02}_{-0.02}$ & -15.13 & -13.82 \\ 
10:00:22.21 & 2:10:19.9 & 462 &  38.63 & 7.6 & 6.5 &  5.1 & -6.00 & 188.71 & 1.70 & -0.00 &  1T & $13.42^{}_{}$ &  $-14.89^{+0.08}_{-0.08}$ & -14.57 & -14.40 \\ 
9:58:56.03 & 2:30:40.9 & 516 &  41.33 & 8.2 & 7.1 &  5.0 & -6.00 & 91.61 & 1.01 & -1.13 &  1T & $0.76^{+1.02}_{-0.31}$ &  $-14.27^{+0.08}_{-0.08}$ & -15.38 & -14.24 \\ 
10:00:09.81 & 2:23:49.9 & 527 &  250.56 & 17.5 & 16.4 &  14.3 & -6.00 & 231.76 & 0.91 & -1.30 &  1T & $0.39^{+0.42}_{-0.29}$ &  $-13.87^{+0.03}_{-0.03}$ & -15.89 & -13.86 \\ 
10:01:41.57 & 2:07:59.4 & 537 &  321.81 & 19.4 & 18.4 &  16.6 & -6.00 & 137.65 & 1.00 & -10 &  1T & $0.65^{}_{}$ &  $-13.67^{+0.03}_{-0.03}$ & -14.98 & -13.65 \\ 
10:01:52.18 & 2:11:58.4 & 546 &  1700.06 & 42.3 & 41.3 &  40.2 & -6.00 & 90.37 & 1.06 & -10 &  1T & $1.01^{}_{}$ &  $-12.85^{+0.01}_{-0.01}$ & -13.64 & -12.79 \\ 
9:59:15.68 & 2:32:25.0 & 578 &  9.35 & 4.4 & 3.3 &  2.1 & -5.47 & 93.49 & 0.96 & -0.35 &  \ldots & \dots &  $-15.13^{}_{}$ & \ldots & \ldots \\ 
10:01:43.18 & 2:17:28.4 & 590 &  392.42 & 21.1 & 20.1 &  18.6 & -6.00 & 181.79 & 0.93 & -0.31 &  1T & $0.57^{}_{}$ &  $-13.71^{+0.02}_{-0.02}$ & -15.19 & -13.69 \\ 
10:00:05.61 & 2:07:00.9 & 742 &  12.95 & 6.1 & 5.0 &  2.1 & -3.03 & 188.61 & 1.28 & -0.84 &  \ldots & \dots &  $-15.11^{}_{}$ & \ldots & \ldots \\ 
10:01:30.74 & 2:06:45.9 & 766 &  4.05 & 5.8 & 4.7 &  0.7 & -0.69 & 186.62 & 4.14 & -0.46 &  \ldots & \dots &  $-15.14^{}_{}$ & \ldots & \ldots \\ 
10:00:46.69 & 2:02:33.4 & 843 &  39.30 & 8.6 & 7.5 &  4.6 & -6.00 & 233.25 & 1.36 & -4.35 &  1T & $0.77^{+3.20}_{-0.48}$ &  $-14.37^{+0.09}_{-0.09}$ & -15.47 & -14.34 \\ 
10:00:52.92 & 1:57:14.1 & 847 &  69.69 & 10.7 & 9.6 &  6.5 & -6.00 & 232.32 & 1.04 & -3.37 &  1T & $0.54^{+0.67}_{-0.44}$ &  $-14.47^{+0.06}_{-0.06}$ & -16.02 & -14.46 \\ 
9:59:55.23 & 2:08:44.7 & 870 &  4.75 & 7.2 & 6.1 &  0.7 & -0.64 & 234.43 & 1.07 & -0.71 &  \ldots & \dots &  $-15.76^{}_{}$ & \ldots & \ldots \\ 
9:59:30.81 & 2:32:39.7 & 904 &  15.23 & 6.5 & 5.4 &  2.4 & -3.51 & 185.82 & 1.00 & -0.78 &  \ldots & \dots &  $-15.14^{}_{}$ & \ldots & \ldots \\ 
10:01:09.03 & 2:13:51.1 & 939 &  15.00 & 6.2 & 5.1 &  2.4 & -3.93 & 234.84 & 1.82 & -0.39 &  \ldots & \dots &  $-14.95^{}_{}$ & \ldots & \ldots \\ 
10:00:37.03 & 2:26:14.8 & 989 &  42.55 & 8.1 & 7.0 &  5.3 & -6.00 & 187.36 & 1.09 & -4.49 &  1T & $0.56^{+0.83}_{-0.37}$ &  $-14.40^{+0.08}_{-0.08}$ & -15.91 & -14.39 \\ 
9:59:01.12 & 1:57:38.9 & 998 &  10.16 & 5.7 & 4.6 &  1.8 & -2.32 & 189.19 & 2.04 & -0.56 &  \ldots & \dots &  $-15.06^{}_{}$ & \ldots & \ldots \\ 
9:59:00.98 & 2:08:30.6 & 1056 &  73.97 & 10.1 & 9.1 &  7.3 & -6.00 & 185.90 & 1.10 & -4.23 &  1T & $1.27^{}_{}$ &  $-14.54^{+0.06}_{-0.06}$ & -15.09 & -14.43 \\ 
9:59:29.44 & 2:05:13.5 & 1073 &  17.31 & 6.3 & 5.2 &  2.8 & -5.21 & 189.34 & 1.00 & -0.33 &  \ldots & \dots &  $-15.16^{}_{}$ & \ldots & \ldots \\ 
10:00:55.18 & 1:59:37.6 & 1103 &  24.52 & 7.9 & 6.8 &  3.1 & -5.33 & 232.32 & 1.06 & -0.64 &  1T & $0.21^{}_{}$ &  $-13.55^{+0.12}_{-0.14}$ & -17.11 & -13.55 \\ 
9:59:41.82 & 2:08:59.6 & 1137 &  11.51 & 6.5 & 5.4 &  1.8 & -2.08 & 234.31 & 3.26 & -2.80 &  \ldots & \dots &  $-14.86^{}_{}$ & \ldots & \ldots \\ 
10:00:33.51 & 2:05:43.6 & 1173 &  29.82 & 8.4 & 7.4 &  3.5 & -6.00 & 234.75 & 1.06 & -0.09 &  1T & $0.54^{+0.85}_{-0.28}$ &  $-14.63^{+0.11}_{-0.12}$ & -16.19 & -14.61 \\ 
10:02:07.84 & 2:22:34.9 & 1560 &  29.23 & 6.8 & 5.7 &  4.3 & -6.00 & 89.31 & 0.90 & -0.64 &  1T & $0.36^{+0.64}_{-0.09}$ &  $-14.30^{+0.09}_{-0.09}$ & -16.50 & -14.30 \\ 
10:01:43.23 & 2:32:52.8 & 1592 &  81.57 & 10.9 & 9.9 &  7.5 & -6.00 & 89.30 & 0.97 & -0.73 &  1T & $0.67^{}_{}$ &  $-13.94^{+0.05}_{-0.06}$ & -15.21 & -13.92 \\ 
9:59:11.16 & 2:42:24.0 & 1600 &  5.26 & 4.3 & 3.1 &  1.2 & -1.58 & 46.67 & 3.79 & -0.47 &  \ldots & \dots &  $-14.43^{}_{}$ & \ldots & \ldots \\ 
9:59:18.33 & 2:43:05.2 & 1604 &  32.37 & 7.1 & 6.0 &  4.6 & -6.00 & 46.67 & 1.22 & -2.33 &  1T & $1.36^{}_{}$ &  $-14.28^{+0.09}_{-0.09}$ & -14.77 & -14.16 \\ 
9:58:04.42 & 1:52:16.8 & 1688 &  41.96 & 8.6 & 7.5 &  4.9 & -6.00 & 48.26 & 1.22 & -1.58 &  1T & $1.65^{}_{}$ &  $-14.18^{+0.08}_{-0.09}$ & -14.53 & -14.02 \\ 
9:59:08.27 & 1:57:32.9 & 1710 &  2.65 & 6.7 & 5.6 &  0.4 & -0.47 & 285.04 & 0.81 & -2.22 &  \ldots & \dots &  $-16.17^{}_{}$ & \ldots & \ldots \\ 
10:02:21.95 & 2:20:41.9 & 1768 &  4.39 & 5.0 & 3.9 &  0.9 & -0.90 & 89.79 & 0.78 & -0.25 &  \ldots & \dots &  $-15.51^{}_{}$ & \ldots & \ldots \\ 
10:01:16.77 & 2:17:13.9 & 2061 &  38.36 & 10.7 & 9.7 &  3.6 & -5.71 & 369.12 & 1.61 & -4.44 &  1T & $5.92^{}_{}$ &  $-15.18^{+0.11}_{-0.13}$ & -15.00 & -14.78 \\ 
9:59:06.13 & 2:34:11.1 & 2216 &  16.13 & 6.0 & 4.9 &  2.7 & -5.30 & 93.49 & 1.41 & -1.00 &  \ldots & \dots &  $-14.72^{}_{}$ & \ldots & \ldots \\ 
10:02:01.70 & 2:03:55.5 & 2331 &  9.24 & 5.6 & 4.5 &  1.6 & -2.06 & 92.44 & 0.97 & -0.06 &  \ldots & \dots &  $-15.07^{}_{}$ & \ldots & \ldots \\ 
9:59:10.23 & 2:23:34.8 & 2524 &  6.19 & 4.0 & 2.8 &  1.6 & -3.06 & 91.61 & 0.78 & -0.02 &  \ldots & \dots &  $-15.38^{}_{}$ & \ldots & \ldots \\ 
9:59:17.54 & 2:22:06.7 & 2539 &  20.98 & 7.2 & 6.1 &  2.9 & -5.09 & 182.73 & 0.91 & -1.85 &  1T & $0.27^{+0.46}_{-0.11}$ &  $-14.34^{+0.13}_{-0.15}$ & -17.19 & -14.34 \\ 
9:59:02.31 & 2:15:20.3 & 2881 &  6.95 & 6.0 & 4.9 &  1.1 & -1.16 & 133.87 & 1.06 & -3.34 &  \ldots & \dots &  $-15.38^{}_{}$ & \ldots & \ldots \\ 
9:58:08.76 & 2:00:01.1 & 3205 &  11.58 & 6.7 & 5.6 &  1.7 & -1.97 & 95.61 & 1.63 & -0.06 &  \ldots & \dots &  $-14.63^{}_{}$ & \ldots & \ldots \\ 
9:58:39.08 & 2:09:05.8 & 3232 &  13.82 & 5.2 & 4.1 &  2.6 & -6.00 & 91.46 & 0.88 & -0.20 &  \ldots & \dots &  $-15.01^{}_{}$ & \ldots & \ldots \\ 
9:58:51.21 & 2:02:26.8 & 3243 &  12.59 & 5.7 & 4.6 &  2.2 & -3.59 & 189.19 & 0.91 & -0.33 &  \ldots & \dots &  $-15.33^{}_{}$ & \ldots & \ldots \\ 
10:00:45.93 & 1:48:19.9 & 3353 &  5.08 & 5.3 & 4.1 &  1.0 & -0.99 & 186.67 & 0.85 & -0.30 &  \ldots & \dots &  $-15.73^{}_{}$ & \ldots & \ldots \\ 
10:00:03.59 & 1:50:44.9 & 3381 &  23.60 & 8.0 & 6.9 &  3.0 & -4.76 & 237.81 & 0.87 & -3.26 &  1T & $0.43^{}_{}$ &  $-14.91^{+0.13}_{-0.15}$ & -16.79 & -14.91 \\ 
9:59:20.91 & 1:52:03.6 & 3425 &  6.96 & 4.0 & 2.8 &  1.8 & -4.53 & 93.67 & 1.00 & -0.66 &  \ldots & \dots &  $-15.18^{}_{}$ & \ldots & \ldots \\ 
9:59:39.21 & 1:53:49.8 & 3452 &  12.88 & 7.6 & 6.5 &  1.7 & -1.79 & 237.30 & 1.04 & -2.30 &  \ldots & \dots &  $-15.24^{}_{}$ & \ldots & \ldots \\ 
9:59:12.91 & 2:00:58.4 & 3517 &  17.05 & 8.0 & 6.9 &  2.1 & -2.55 & 236.78 & 2.20 & -2.19 &  \ldots & \dots &  $-14.80^{}_{}$ & \ldots & \ldots \\ 
10:00:40.34 & 2:36:56.2 & 3650 &  77.04 & 10.6 & 9.5 &  7.3 & -6.00 & 94.46 & 1.12 & -2.40 &  1T & $1.71^{}_{}$ &  $-14.21^{+0.06}_{-0.06}$ & -14.53 & -14.04 \\ 
10:00:55.31 & 2:33:30.4 & 3664 &  19.90 & 5.8 & 4.7 &  3.4 & -6.00 & 92.79 & 1.06 & -0.17 &  \ldots & \dots &  $-14.65^{}_{}$ & \ldots & \ldots \\ 
10:00:36.92 & 2:23:57.5 & 3683 &  8.78 & 5.2 & 4.1 &  1.7 & -2.24 & 187.36 & 0.96 & -0.03 &  \ldots & \dots &  $-15.45^{}_{}$ & \ldots & \ldots \\ 
10:01:18.22 & 2:05:52.4 & 3782 &  3.25 & 4.8 & 3.6 &  0.7 & -0.72 & 186.62 & 2.23 & -0.47 &  \ldots & \dots &  $-15.49^{}_{}$ & \ldots & \ldots \\ 
9:59:50.63 & 2:23:15.9 & 3811 &  50.30 & 9.9 & 8.9 &  5.1 & -6.00 & 277.19 & 1.26 & -5.22 &  1T & $0.24^{+1.15}_{-0.15}$ &  $-13.23^{+0.08}_{-0.08}$ & -16.35 & -13.23 \\ 
9:59:10.21 & 1:53:14.2 & 10552 &  3.86 & 3.8 & 2.6 &  1.0 & -1.34 & 93.67 & 0.96 & -0.08 &  \ldots & \dots &  $-15.48^{}_{}$ & \ldots & \ldots \\ 
10:00:11.46 & 2:28:34.0 & 10742 &  16.36 & 5.8 & 4.7 &  2.8 & -6.00 & 186.93 & 0.96 & -0.42 &  \ldots & \dots &  $-15.16^{}_{}$ & \ldots & \ldots \\ 
10:01:35.76 & 2:03:34.7 & 11145 &  11.00 & 5.2 & 4.1 &  2.1 & -3.60 & 140.82 & 0.81 & -0.33 &  \ldots & \dots &  $-15.24^{}_{}$ & \ldots & \ldots \\ 
10:00:54.50 & 2:16:05.1 & 11537 &  3.48 & 5.0 & 3.9 &  0.7 & -0.71 & 189.69 & 1.03 & -1.56 &  \ldots & \dots &  $-15.83^{}_{}$ & \ldots & \ldots \\ 
10:01:28.50 & 1:59:32.4 & 11905 &  4.15 & 5.0 & 3.9 &  0.8 & -0.85 & 184.20 & 0.87 & -1.07 &  \ldots & \dots &  $-15.61^{}_{}$ & \ldots & \ldots \\ 
10:01:02.45 & 2:22:29.7 & 12635 &  8.00 & 5.4 & 4.2 &  1.5 & -1.79 & 184.79 & 1.00 & -3.29 &  \ldots & \dots &  $-15.49^{}_{}$ & \ldots & \ldots \\ 
\label{xstars_cosmos} 
\end{longtable} 
Notes. Columns 1-2: Source position (from optical images). Column 3: {\it Chandra} COSMOS ID. Column 4: Net counts in the full (0.5-8.0 keV) band. Columns 5-6: Upper and lower 1$\sigma$ errors on the number of net counts. Column 7: Source detection significance. Column 8: logarithm of the Poisson probability that the source is a chance coincidence of background events. Values below -6.0 are listed as -6.0. Column 9: Full exposure time for each source derived from the mono-energetic exposure maps for the combined observations. Column 10: Background corrected median energy of all source photons in the full (0.5-8.0 keV) band. Column 11: Logarithm of the Kolmogorov-Smirnov probability that the source is not variable. Column 12: X-ray spectral model fit type: single-temperature thermal plasma model (1T) or no model fit (-). Column 13: Thermal plasma temperature of model fit with upper and lower 90\% confidence intervals (uncertainties missing when they are so large that the parameter is effectively unconstrained). Column 14: Logarithm of the extinction-corrected X-ray flux in the full (0.5-8.0 keV) band from model fit or derived from the number of net counts for unfit sources as described in Section~2. Upper and lower 1$\sigma$ errors are shown or left blank when the upper or lower bounds are unconstrained. For sources without model fits the flux errors are not specificed individually. Columns 15-16: Logarithm of extinction-corrected soft (0.5-2.0 keV) and hard (2.0-8.0 keV) band fluxes.
\clearpage 
\end{landscape}

\tiny 
\begin{landscape}
\begin{longtable}{ccc ccccc p{0.001cm} ccc p{0.001cm} ccc}
\caption[]{Optical and near-IR properties of stellar sources detected in the {\it Chandra} COSMOS survey.}\\ 
\tableline 
RA & Dec & CID & \multicolumn{5}{c}{Optical photometry} && \multicolumn{3}{c}{Near-IR photometry} && \multicolumn{3}{c}{Spectral information} \\ 
\cline{4-8} \cline{10-12} \cline{14-16} 
 (J2000) & (J2000) & & $u$ & $g$ & $r$ & $i$ & $z$ && $J$ & $H$ & $K_s$ && Type & Orig. & d (kpc) \\ 
\tableline 
\endhead 
\tableline 
\multicolumn{16}{r}{\emph{continued on next page}}\\ 
\endfoot 
\tableline 
\endlastfoot 
10:00:48.44 & 2:07:34.8 & 48 &  &  &  &  &  && 11.850 $\pm$ 0.023 & 11.517 $\pm$ 0.023 & 11.447 $\pm$ 0.023 && G7$^{}$ & F & 0.35 \\ 
10:00:49.51 & 2:07:14.6 & 49 & 22.1241 $\pm$ 0.1166 & 20.099 $\pm$ 0.011 & 18.698 $\pm$ 0.006 & 17.096 $\pm$ 0.003 & 16.249 $\pm$ 0.004 && 14.762 $\pm$ 0.038 & 14.255 $\pm$ 0.033 & 13.864 $\pm$ 0.058 && M5$^{}$ & V & 0.14 \\ 
10:00:20.96 & 1:44:32.3 & 268 & 23.1336 $\pm$ 0.2585 & 19.871 $\pm$ 0.010 & 18.285 $\pm$ 0.004 & 16.655 $\pm$ 0.002 & 15.785 $\pm$ 0.002 && 14.251 $\pm$ 0.035 & 13.756 $\pm$ 0.034 & 13.495 $\pm$ 0.047 && M5$^{}$ & VI & 0.12 \\ 
9:58:23.06 & 2:13:11.9 & 321 &  &  &  &  &  && 10.397 $\pm$ 0.024 & 10.162 $\pm$ 0.027 & 10.092 $\pm$ 0.023 && F8$^{}$ & F & 0.26 \\ 
10:00:57.46 & 1:55:48.7 & 367 &  &  &  &  &  && 10.755 $\pm$ 0.023 & 10.566 $\pm$ 0.027 & 10.519 $\pm$ 0.023 && F5$^{}$ & F & 0.38 \\ 
10:01:42.18 & 1:53:19.7 & 397 &  &  &  &  &  && 9.946 $\pm$ 0.024 & 9.548 $\pm$ 0.025 & 9.435 $\pm$ 0.024 && G9$^{}$ & F & 0.13 \\ 
9:59:54.70 & 2:17:06.0 & 444 & 18.2721 $\pm$ 0.004 & 15.48 $\pm$ 0.001 & 14.069 $\pm$ 0.001 & 14.928 $\pm$ 0.001 & 13.014 $\pm$ 0.000 && 11.681 $\pm$ 0.024 & 11.020 $\pm$ 0.022 & 10.845 $\pm$ 0.020 && M1$^{}$ & F & 0.12 \\ 
10:00:22.21 & 2:10:19.9 & 462 & 22.8733 $\pm$ 0.2368 & 20.404 $\pm$ 0.015 & 19.002 $\pm$ 0.007 & 18.23 $\pm$ 0.005 & 17.79 $\pm$ 0.017 && 16.642 $\pm$ 0.002 &  & 15.853 $\pm$ 0.003 && M1$^{}$ & p & 1.19 \\ 
9:58:56.03 & 2:30:40.9 & 516 & 20.7987 $\pm$ 0.0234 & 17.857 $\pm$ 0.003 & 16.502 $\pm$ 0.002 & 15.918 $\pm$ 0.001 & 15.58 $\pm$ 0.002 && 13.702 $\pm$ 0.033 & 13.085 $\pm$ 0.033 & 12.884 $\pm$ 0.034 && K7$^{}$ & F & 0.43 \\ 
10:00:09.81 & 2:23:49.9 & 527 &  &  &  &  &  && 8.766 $\pm$ 0.025 & 8.215 $\pm$ 0.031 & 8.159 $\pm$ 0.031 && K2$^{}$ & F & 0.06 \\ 
10:01:41.57 & 2:07:59.4 & 537 & 20.3912 $\pm$ 0.0239 & 17.637 $\pm$ 0.003 & 16.152 $\pm$ 0.001 & 14.826 $\pm$ 0.001 & 14.118 $\pm$ 0.001 && 12.693 $\pm$ 0.026 & 12.106 $\pm$ 0.025 & 11.827 $\pm$ 0.026 && M3e$^{}$ & F & 0.12 \\ 
10:01:52.18 & 2:11:58.4 & 546 &  &  &  &  &  && 11.069 $\pm$ 0.026 & 10.451 $\pm$ 0.024 & 10.318 $\pm$ 0.021 && K7e$^{}$ & F & 0.13 \\ 
9:59:15.68 & 2:32:25.0 & 578 & 20.6441 $\pm$ 0.0265 & 17.754 $\pm$ 0.003 & 16.254 $\pm$ 0.001 & 15.329 $\pm$ 0.001 & 14.717 $\pm$ 0.001 && 13.545 $\pm$ 0.027 & 12.909 $\pm$ 0.026 & 12.692 $\pm$ 0.030 && M2$^{}$ & F & 0.23 \\ 
10:01:43.18 & 2:17:28.4 & 590 &  &  &  &  &  && 7.038 $\pm$ 0.017 & 6.567 $\pm$ 0.029 & 6.461 $\pm$ 0.024 && G7$^{}$ & F & 0.04 \\ 
10:00:05.61 & 2:07:00.9 & 742 &  & 25.133 $\pm$ 0.118 & 23.447 $\pm$ 0.037 & 21.741 $\pm$ 0.108 & 20.45 $\pm$ 0.012 && 18.782 $\pm$ 0.005 &  & 17.992 $\pm$ 0.009 && M6$^{}$ & p & 0.58 \\ 
10:01:30.74 & 2:06:45.9 & 766 & 17.1379 $\pm$ 0.002 & 15.754 $\pm$ 0.001 & 15.428 $\pm$ 0.001 & 15.388 $\pm$ 0.001 & 15.326 $\pm$ 0.002 && 14.468 $\pm$ 0.038 & 14.065 $\pm$ 0.048 & 14.011 $\pm$ 0.071 && F7$^{}$ & F & 1.85 \\ 
10:00:46.69 & 2:02:33.4 & 843 & 23.2358 $\pm$ 0.1812 & 21.395 $\pm$ 0.031 & 19.945 $\pm$ 0.015 & 18.252 $\pm$ 0.005 & 17.316 $\pm$ 0.007 && 15.806 $\pm$ 0.001 & 15.168 $\pm$ 0.095 & 15.052 $\pm$ 0.002 && M6$^{}$ & V & 0.15 \\ 
10:00:52.92 & 1:57:14.1 & 847 & 24.1586 $\pm$ 0.4215 & 22.598 $\pm$ 0.024 & 20.894 $\pm$ 0.032 & 18.637 $\pm$ 0.007 & 17.382 $\pm$ 0.007 && 15.705 $\pm$ 0.001 & 15.207 $\pm$ 0.107 & 14.773 $\pm$ 0.001 && M7$^{}$ & V & 0.10 \\ 
9:59:55.23 & 2:08:44.7 & 870 & 18.7864 $\pm$ 0.0067 & 15.796 $\pm$ 0.001 & 14.428 $\pm$ 0.001 & 13.897 $\pm$ 0.000 & 13.538 $\pm$ 0.001 && 12.223 $\pm$ 0.033 & 11.562 $\pm$ 0.035 & 11.397 $\pm$ 0.029 && M0$^{}$ & p & 0.17 \\ 
9:59:30.81 & 2:32:39.7 & 904 & 20.6471 $\pm$ 0.0308 & 17.768 $\pm$ 0.003 & 16.273 $\pm$ 0.001 & 15.453 $\pm$ 0.001 & 15.034 $\pm$ 0.001 && 13.821 $\pm$ 0.024 & 13.139 $\pm$ 0.022 & 12.931 $\pm$ 0.034 && M1$^{}$ & F & 0.31 \\ 
10:01:09.03 & 2:13:51.1 & 939 & 21.964 $\pm$ 0.1007 & 19.218 $\pm$ 0.006 & 18.223 $\pm$ 0.004 & 17.837 $\pm$ 0.004 & 17.699 $\pm$ 0.014 && 16.677 $\pm$ 0.002 &  & 15.908 $\pm$ 0.003 && K7$^{}$ & V & 1.74 \\ 
10:00:37.03 & 2:26:14.8 & 989 & 20.3733 $\pm$ 0.0216 & 17.639 $\pm$ 0.003 & 16.481 $\pm$ 0.002 & 16.042 $\pm$ 0.001 & 15.795 $\pm$ 0.002 && 14.690 $\pm$ 0.041 & 13.974 $\pm$ 0.038 & 13.910 $\pm$ 0.061 && K7$^{}$ & F & 0.69 \\ 
9:59:01.12 & 1:57:38.9 & 998 &  &  &  &  &  && 11.418 $\pm$ 0.023 & 11.139 $\pm$ 0.021 & 11.059 $\pm$ 0.023 && F8$^{}$ & F & 0.41 \\ 
9:59:00.98 & 2:08:30.6 & 1056 & 22.0476 $\pm$ 0.1239 & 19.421 $\pm$ 0.007 & 17.995 $\pm$ 0.004 & 17.079 $\pm$ 0.003 & 16.559 $\pm$ 0.005 && 15.414 $\pm$ 0.062 & 14.755 $\pm$ 0.060 & 14.496 $\pm$ 0.095 && M1e$^{}$ & VI & 0.64 \\ 
9:59:29.44 & 2:05:13.5 & 1073 &  &  &  &  &  && 12.861 $\pm$ 0.024 & 12.551 $\pm$ 0.026 & 12.503 $\pm$ 0.021 && F6$^{}$ & F & 0.90 \\ 
10:00:55.18 & 1:59:37.6 & 1103 & 23.0389 $\pm$ 0.1515 & 20.635 $\pm$ 0.017 & 19.216 $\pm$ 0.008 & 17.851 $\pm$ 0.004 & 17.115 $\pm$ 0.006 && 15.746 $\pm$ 0.001 & 15.147 $\pm$ 0.097 & 14.824 $\pm$ 0.001 && M4e$^{\ast}$ & V & 0.33 \\ 
9:59:41.82 & 2:08:59.6 & 1137 & 23.4928 $\pm$ 0.4421 & 21.884 $\pm$ 0.017 & 20.454 $\pm$ 0.022 & 19.404 $\pm$ 0.014 & 18.889 $\pm$ 0.040 && 17.649 $\pm$ 0.003 &  & 16.808 $\pm$ 0.004 && M2$^{}$ & p & 1.55 \\ 
10:00:33.51 & 2:05:43.6 & 1173 & 22.8809 $\pm$ 0.2385 & 19.637 $\pm$ 0.009 & 18.162 $\pm$ 0.004 & 16.792 $\pm$ 0.002 & 16.048 $\pm$ 0.004 && 14.655 $\pm$ 0.035 & 14.012 $\pm$ 0.042 & 13.638 $\pm$ 0.048 && M3$^{}$ & I & 0.28 \\ 
10:02:07.84 & 2:22:34.9 & 1560 & 19.796 $\pm$ 0.0111 & 16.975 $\pm$ 0.002 & 15.523 $\pm$ 0.001 & 15.628 $\pm$ 0.001 & 13.463 $\pm$ 0.000 && 12.083 $\pm$ 0.024 & 11.481 $\pm$ 0.024 & 11.220 $\pm$ 0.026 && M3$^{}$ & F & 0.09 \\ 
10:01:43.23 & 2:32:52.8 & 1592 & 19.9122 $\pm$ 0.0153 & 17.121 $\pm$ 0.002 & 15.824 $\pm$ 0.001 & 15.138 $\pm$ 0.001 & 14.757 $\pm$ 0.001 && 13.306 $\pm$ 0.026 & 12.639 $\pm$ 0.022 & 12.458 $\pm$ 0.024 && K7e$^{}$ & I & 0.36 \\ 
9:59:11.16 & 2:42:24.0 & 1600 & 23.8312 $\pm$ 0.3835 & 22.058 $\pm$ 0.019 & 21.356 $\pm$ 0.013 & 21.193 $\pm$ 0.071 & 20.864 $\pm$ 0.196 && 19.916 $\pm$ 0.019 &  & 19.467 $\pm$ 0.040 && K1$^{}$ & p & 11.73 \\ 
9:59:18.33 & 2:43:05.2 & 1604 & 21.2441 $\pm$ 0.0368 & 18.305 $\pm$ 0.004 & 17.003 $\pm$ 0.002 & 16.262 $\pm$ 0.002 & 15.774 $\pm$ 0.002 && 14.435 $\pm$ 0.033 & 13.834 $\pm$ 0.027 & 13.714 $\pm$ 0.047 && M0e$^{}$ & I & 0.51 \\ 
9:58:04.42 & 1:52:16.8 & 1688 & 19.8183 $\pm$ 0.0097 & 17.209 $\pm$ 0.002 & 16.072 $\pm$ 0.001 & 15.571 $\pm$ 0.001 & 15.276 $\pm$ 0.001 && 13.938 $\pm$ 0.023 & 13.308 $\pm$ 0.029 & 13.125 $\pm$ 0.034 && K7$^{}$ & F & 0.48 \\ 
9:59:08.27 & 1:57:32.9 & 1710 &  &  &  &  &  && 10.542 $\pm$ 0.026 & 10.138 $\pm$ 0.029 & 10.097 $\pm$ 0.026 && G7$^{}$ & F & 0.19 \\ 
10:02:21.95 & 2:20:41.9 & 1768 & 23.3565 $\pm$ 0.2846 & 21.463 $\pm$ 0.033 & 20.147 $\pm$ 0.015 & 18.794 $\pm$ 0.009 & 18.087 $\pm$ 0.018 && 16.678 $\pm$ 0.002 &  & 15.829 $\pm$ 0.002 && M3$^{}$ & p & 0.76 \\ 
10:01:16.77 & 2:17:13.9 & 2061 & 19.6953 $\pm$ 0.0103 & 17.405 $\pm$ 0.002 & 16.484 $\pm$ 0.002 & 16.102 $\pm$ 0.001 & 15.917 $\pm$ 0.002 && 14.667 $\pm$ 0.060 & 14.168 $\pm$ 0.065 & 13.982 $\pm$ 0.089 && K3$^{}$ & F & 0.85 \\ 
9:59:06.13 & 2:34:11.1 & 2216 & 17.7816 $\pm$ 0.0028 & 15.499 $\pm$ 0.001 & 15.134 $\pm$ 0.001 & 14.339 $\pm$ 0.001 & 14.156 $\pm$ 0.001 && 12.874 $\pm$ 0.026 & 12.270 $\pm$ 0.026 & 12.142 $\pm$ 0.026 && G8$^{}$ & F & 0.84 \\ 
10:02:01.70 & 2:03:55.5 & 2331 &  &  &  &  &  && 11.926 $\pm$ 0.024 & 11.319 $\pm$ 0.023 & 11.226 $\pm$ 0.023 && K5$^{}$ & F & 0.22 \\ 
9:59:10.23 & 2:23:34.8 & 2524 & 17.1923 $\pm$ 0.0017 & 15.122 $\pm$ 0.001 & 14.49 $\pm$ 0.001 & 15.224 $\pm$ 0.001 & 14.155 $\pm$ 0.001 && 12.982 $\pm$ 0.027 & 12.550 $\pm$ 0.023 & 12.559 $\pm$ 0.021 && K2$^{}$ & F & 0.46 \\ 
9:59:17.54 & 2:22:06.7 & 2539 &  &  &  &  &  && 10.882 $\pm$ 0.024 & 10.381 $\pm$ 0.021 & 10.267 $\pm$ 0.021 && K3$^{}$ & F & 0.15 \\ 
9:59:02.31 & 2:15:20.3 & 2881 & 24.2583 $\pm$ 0.4817 & 23.331 $\pm$ 0.034 & 21.543 $\pm$ 0.013 & 19.956 $\pm$ 0.016 & 18.808 $\pm$ 0.025 && 17.335 $\pm$ 0.002 &  & 16.509 $\pm$ 0.004 && M6$^{}$ & p & 0.29 \\ 
9:58:08.76 & 2:00:01.1 & 3205 & 20.5412 $\pm$ 0.0174 & 19.219 $\pm$ 0.006 & 18.81 $\pm$ 0.006 & 18.684 $\pm$ 0.007 & 18.656 $\pm$ 0.023 && 17.767 $\pm$ 0.004 &  & 17.311 $\pm$ 0.006 && F8$^{}$ & p & 8.02 \\ 
9:58:39.08 & 2:09:05.8 & 3232 &  &  &  &  &  && 10.398 $\pm$ 0.022 & 9.785 $\pm$ 0.026 & 9.658 $\pm$ 0.023 && K7$^{}$ & F & 0.10 \\ 
9:58:51.21 & 2:02:26.8 & 3243 &  &  &  &  &  && 12.138 $\pm$ 0.023 & 11.636 $\pm$ 0.023 & 11.513 $\pm$ 0.023 && K3$^{}$ & F & 0.27 \\ 
10:00:45.93 & 1:48:19.9 & 3353 & 21.6116 $\pm$ 0.0824 & 18.509 $\pm$ 0.004 & 17.117 $\pm$ 0.002 & 16.301 $\pm$ 0.002 & 15.897 $\pm$ 0.002 && 14.579 $\pm$ 0.037 & 13.973 $\pm$ 0.042 & 13.708 $\pm$ 0.054 && M1e$^{}$ & F & 0.44 \\ 
10:00:03.59 & 1:50:44.9 & 3381 &  &  &  &  &  && 9.219 $\pm$ 0.035 & 8.927 $\pm$ 0.049 & 8.721 $\pm$ 0.033 && G8$^{}$ & F & 0.09 \\ 
9:59:20.91 & 1:52:03.6 & 3425 & 24.5451 $\pm$ 0.6176 & 22.055 $\pm$ 0.018 & 20.574 $\pm$ 0.022 & 19.501 $\pm$ 0.014 & 18.927 $\pm$ 0.026 && 17.640 $\pm$ 0.003 &  & 16.844 $\pm$ 0.004 && M3$^{}$ & p & 1.20 \\ 
9:59:39.21 & 1:53:49.8 & 3452 & 24.1914 $\pm$ 0.4467 & 22.671 $\pm$ 0.025 & 21.456 $\pm$ 0.048 & 19.551 $\pm$ 0.015 & 18.705 $\pm$ 0.022 && 17.095 $\pm$ 0.003 &  & 16.268 $\pm$ 0.003 && M5$^{}$ & p & 0.42 \\ 
9:59:12.91 & 2:00:58.4 & 3517 & 21.2677 $\pm$ 0.0238 & 18.551 $\pm$ 0.004 & 17.372 $\pm$ 0.002 & 16.905 $\pm$ 0.002 & 16.653 $\pm$ 0.003 && 15.556 $\pm$ 0.001 & 14.804 $\pm$ 0.055 & 14.726 $\pm$ 0.001 && K5$^{}$ & F & 1.10 \\ 
10:00:40.34 & 2:36:56.2 & 3650 & 22.5847 $\pm$ 0.1697 & 19.864 $\pm$ 0.010 & 18.411 $\pm$ 0.005 & 16.833 $\pm$ 0.002 & 15.979 $\pm$ 0.002 && 14.563 $\pm$ 0.036 & 13.903 $\pm$ 0.043 & 13.668 $\pm$ 0.054 && M5$^{}$ & VI & 0.13 \\ 
10:00:55.31 & 2:33:30.4 & 3664 & 23.3307 $\pm$ 0.3363 & 21.114 $\pm$ 0.027 & 19.628 $\pm$ 0.012 & 17.785 $\pm$ 0.004 & 16.804 $\pm$ 0.005 && 15.240 $\pm$ 0.048 & 14.646 $\pm$ 0.069 & 14.341 $\pm$ 0.084 && M6e$^{}$ & I & 0.11 \\ 
10:00:36.92 & 2:23:57.5 & 3683 &  &  &  &  &  && 11.264 $\pm$ 0.021 & 10.833 $\pm$ 0.024 & 10.776 $\pm$ 0.025 && G9$^{}$ & F & 0.23 \\ 
10:01:18.22 & 2:05:52.4 & 3782 &  &  &  &  &  && 11.617 $\pm$ 0.025 & 11.181 $\pm$ 0.026 & 11.092 $\pm$ 0.027 && K2$^{}$ & F & 0.23 \\ 
9:59:50.63 & 2:23:15.9 & 3811 & 20.8651 $\pm$ 0.033 & 18.614 $\pm$ 0.004 & 17.734 $\pm$ 0.003 & 17.328 $\pm$ 0.003 & 17.071 $\pm$ 0.007 && 16.003 $\pm$ 0.001 & 15.546 $\pm$ 0.082 & 15.261 $\pm$ 0.002 && K7$^{}$ & I & 1.29 \\ 
9:59:10.21 & 1:53:14.2 & 10552 & 19.6035 $\pm$ 0.0082 & 17.688 $\pm$ 0.003 & 17.029 $\pm$ 0.002 & 16.762 $\pm$ 0.002 & 16.655 $\pm$ 0.004 && 15.655 $\pm$ 0.001 & 15.002 $\pm$ 0.068 & 15.028 $\pm$ 0.002 && K3$^{}$ & F & 1.38 \\ 
10:00:11.46 & 2:28:34.0 & 10742 & 21.5494 $\pm$ 0.0461 & 18.618 $\pm$ 0.004 & 17.302 $\pm$ 0.002 & 16.676 $\pm$ 0.002 & 16.33 $\pm$ 0.003 && 15.175 $\pm$ 0.061 & 14.616 $\pm$ 0.056 & 14.354 $\pm$ 0.100 && K7$^{}$ & F & 0.85 \\ 
10:01:35.76 & 2:03:34.7 & 11145 & 16.4887 $\pm$ 0.0012 & 14.076 $\pm$ 0.001 & 14.547 $\pm$ 0.001 & 14.641 $\pm$ 0.001 & 12.775 $\pm$ 0.000 && 11.828 $\pm$ 0.023 & 11.285 $\pm$ 0.023 & 11.207 $\pm$ 0.025 && K3$^{}$ & F & 0.24 \\ 
10:00:54.50 & 2:16:05.1 & 11537 & 23.4609 $\pm$ 0.2639 & 21.355 $\pm$ 0.030 & 19.888 $\pm$ 0.010 & 18.279 $\pm$ 0.005 & 17.394 $\pm$ 0.007 && 15.964 $\pm$ 0.001 & 15.292 $\pm$ 0.104 & 15.058 $\pm$ 0.002 && M5$^{}$ & p & 0.24 \\ 
10:01:28.50 & 1:59:32.4 & 11905 & 21.8892 $\pm$ 0.0538 & 19.987 $\pm$ 0.010 & 18.725 $\pm$ 0.006 & 17.597 $\pm$ 0.003 & 16.858 $\pm$ 0.004 && 15.501 $\pm$ 0.001 & 14.906 $\pm$ 0.076 & 14.655 $\pm$ 0.001 && M2$^{}$ & p & 0.57 \\ 
10:01:02.45 & 2:22:29.7 & 12635 & 17.1883 $\pm$ 0.0018 & 15.824 $\pm$ 0.001 & 15.41 $\pm$ 0.001 & 15.245 $\pm$ 0.001 & 15.206 $\pm$ 0.001 && 14.322 $\pm$ 0.032 & 13.953 $\pm$ 0.040 & 13.981 $\pm$ 0.063 && F8$^{}$ & F & 1.68 \\ 
\label{xstars_oir} 
\end{longtable} 
Notes. Columns 1-2: Source position (from optical images). Column 3: {\it Chandra} COSMOS ID. Columns 4-8: Optical photometry with errors. Columns 9-11: Near-IR photometry with errors. Column 12: Spectral type (sources with an asterisk have low-quality spectra and are accurate to $\pm2$ subtypes). Column 13: Source of spectral classification: FAST (F), VIMOS (V), IMACS (I), VIMOS+IMACS (VI) or photometric (p). Column 13: Estimated distance based on spectral type.
\clearpage 
\end{landscape}

\end{document}